\documentclass[5p,authoryear]{elsarticle}

\usepackage{hyperref}
\hypersetup{linkbordercolor=1 1 1, pdfborder=0 0 0}
\usepackage[dvipsnames, svgnames, x11names]{xcolor}
\usepackage[edges]{forest}
\usepackage{booktabs}
\usepackage{array}
\usepackage{paralist}
\definecolor{happy}{HTML}{F0E442}
\definecolor{unhappy}{HTML}{9AD0F3}
\usepackage{url}

\journal{Journal of Systems and Software}

\newcommand{\rqone}{What are the experienced consequences of unhappiness among software developers while developing software?}
\newcommand{\rqtwo}{What are the experienced consequences of happiness among software developers while developing software?}

\begin{document}

\begin{frontmatter}

\title{What happens when software developers are (un)happy}

\author[ustutt]{Daniel Graziotin}
\ead{daniel.graziotin@iste.uni-stuttgart.de}
\author[uhel,bth]{Fabian Fagerholm}
\ead{fabian.fagerholm@helsinki.fi}
\author[unibz]{Xiaofeng Wang}
\ead{xiaofeng.wang@unibz.it}
\author[jyu]{Pekka Abrahamsson}
\ead{pekka.abrahamsson@jyu.fi}

\address[ustutt]{Institute of Software Technology, University of Stuttgart, Germany}
\address[uhel]{Department of Computer Science, University of Helsinki, Finland}
\address[bth]{Software Engineering Research Lab Sweden, Blekinge Institute of Technology, Karlskrona, Sweden}
\address[unibz]{Faculty of Computer Science, Free University of Bozen-Bolzano, Italy}
\address[jyu]{Faculty of Information Technology, University of Jyv{\"a}skyl{\"a}, Jyv{\"a}skyl{\"a}, Finland}

\begin{abstract}
The growing literature on affect among software developers mostly reports on the linkage between happiness, software quality, and developer productivity. Understanding happiness and unhappiness in all its components -- positive and negative emotions and moods -- is an attractive and important endeavor. Scholars in industrial and organizational psychology have suggested that understanding happiness and unhappiness could lead to cost-effective ways of enhancing working conditions, job performance, and to limiting the occurrence of psychological disorders. Our comprehension of the consequences of (un)happiness among developers is still too shallow, being mainly expressed in terms of development productivity and software quality. In this paper, we study what happens when developers are happy and unhappy while developing software. Qualitative data analysis of responses given by 317 questionnaire participants identified 42 consequences of unhappiness and 32 of happiness. We found consequences of happiness and unhappiness that are beneficial and detrimental for developers' mental well-being, the software development process, and the produced artifacts. Our classification scheme, available as open data enables new happiness research opportunities of cause-effect type, and it can act as a guideline for practitioners for identifying damaging effects of unhappiness and for fostering happiness on the job.

\vspace{0.5cm}
\paragraph{Note:} the present PDF is the accepted version of 
Graziotin, D., Fagerholm, F., Wang, X., \& Abrahamsson, P. (2018). What happens when software developers are (un)happy . Journal of Systems and Software, 140, 32-47. doi:10.1016/j.jss.2018.02.041.

The final publication is available as openaccess at http://doi.org/10.1016/j.jss.2018.02.041
\end{abstract}


\end{frontmatter}

\section{Introduction}

The idea of flourishing happiness among developers is often promoted by software companies, which knowingly or accidentally attempt to enact the happy-productive worker thesis~\citep{Zelenski2008}. The happiness of all stakeholders involved in software development is an essential element of company success~\citep{Denning2012b}. Recent research within the scope of behavioral software engineering~\citep{Lenberg:2015bj} has highlighted the relationship between software developer happiness and work-related constructs such as performance and productivity~\citep{graziotin2015you,Graziotin2014PEERJ,graziotin2015feelings,Fagerholm2015,Muller2015,Ortu2015a}, quality~\citep{Khan2010,Destefanis2016}, and the social interactions between developers~\citep{Novielli:2015km, Fagerholm2015}. Most of the studies to date have investigated the positive side of happiness.

While happiness, for the individual, is inherently subjective, research shows that it can be studied objectively~\citep{Diener:1999cl, Kahneman:1999ck}. Objective happiness can be construed as the difference between experienced positive affect and experienced negative affect~\citep{Diener:1999cl, Kahneman:1999ck}. Thus, maximizing happiness may be achieved by either maximizing positive experiences or minimizing negative experiences (or both). Studying both happiness and unhappiness is needed to more thoroughly understand the options and opportunities for increasing net happiness.

Focusing on the negative may already be intuitive to many developers. It is a common occurrence that developers share horror stories about their working experience~\citep{Graziotin2014IEEESW}. Managers in the software profession would benefit from greater understanding of the nature and dynamics of unhappiness among developers, and they could take action to prevent dysfunctional responses among employees~\citep{Vecchio:2000ei}. Further understanding of the benefits of limiting negative experiences on the job in general has been called for~\citep{Diener:1999cl}.

We aim to contribute to improved working conditions and quality of life for software developers by broadening the understanding of the (un)happiness of software developers. We are conducting a series of studies using a large-scale quantitative and qualitative survey of software developers. Through those studies, we seek to assess the distribution of (un)happiness in the developer population, the causes of (un)happiness, and what the consequences of those experiences are. The study described in the present article is part of that series, and focuses on the consequences of (un)happiness, i.e., what software developers consider to happen when they experience happiness or unhappiness. We have previously reported on other parts of the study series, first as a poster-preview of results regarding the consequences of unhappiness~\citep{Graziotin:2017} which preceded a workshop paper on the same subject~\citep{Graziotin:2017semotion}. The present article is an extension of the workshop paper. We have reported on the (un)happiness distribution among software developers and the experienced causes of unhappiness in a separate paper~\citep{Graziotin:2017ease}.

The present article extends the previously reported findings on the consequences of unhappiness~\citep{Graziotin:2017semotion} by including consequences of happiness and considering the two together. Specifically, we investigate the following research questions:

\mbox{}
\begin{compactenum}[RQ1]
\item \emph{\rqone}
\item \emph{\rqtwo}
\end{compactenum}
\mbox{}

We report 42 experienced consequences of unhappiness and 32 of happiness that we identified. The consequences concern developers themselves in the form of cognitive and behavioral changes, and external outcomes related to the software development process and artifacts.

\section{Background and Related Work}
\label{sec:related:background_and_related_work}
What is happiness, and how can it be assessed? Intuitively, happiness is related to an individual's sensing of their own affect. In this section, we give a brief overview of the concepts and theory related to affect, emotions, moods, and happiness. We also discuss previous research on happiness and related constructs in the area of software engineering.

\subsection{Theory of Affect and Happiness}

\emph{Affect} is widely agreed to be \emph{``a neurophysiological state that is consciously accessible as a simple, non-reflective feeling that is an integral blend of hedonic (pleasure -- displeasure) and arousal (sleepy -- activated) values''} \citep[p.~147]{Russell2003}. For the purposes of this paper, we use the theory by \citet{Russell2003}, and we thus consider affect to be the atomic unit upon which moods and emotions are constructed. Following several other authors, e.g.~\citet{Fisher2000} and \citet{Khan2010}, we consider \emph{moods} to be prolonged, unattributed affect, while \emph{emotions} are interrelated events concerning a psychological object -- an episodic process, clearly bounded in time, where an individual perceives their own affect.

From a \emph{hedonistic} viewpoint, \textbf{happiness} is a sequence of experiential episodes \citep{Haybron2001} and being happy (unhappy) is associated with frequent experiences of positive (negative) affect~\citep{Diener2010}.\footnote{Alternative views of happiness exist, e.g. Aristoteles' \emph{eudaimonia}: a person is happy because (s)he conducts a satisfactory life full of quality \citep{Haybron2005}. A review of affect theories is given in \citep{Graziotin2015SSE} and the role of the centrality of affect and happiness is discussed in \citep{graziotin2015you}.} Frequent positive (negative) experiential episodes lead to feeling frequent positive (negative) affect, leading to happiness (unhappiness) represented by a positive (negative) \emph{affect balance}~\citep{Diener2010}. Happy individuals are those who experience positive affect more often than negative, and unhappy individuals are those who experience negative affect more often than positive. The affect balance is the difference between experienced negative and positive affect. Therefore, happy individuals have a positive affect balance and unhappy ones have a negative affect balance~\citep{Lyubomirsky2005, Diener2010}.

\subsection{Affect and Happiness in Software Engineering Research}

A considerable increase in the interest of studying affect and happiness among software developers is visible over the last five years, although the research is in its infancy; many theoretical and methodological issues remain in software engineering research, as illustrated by \citet{Graziotin2015SSE,graziotin2015affect} and by \citet{Novielli:2015km}.

Several studies have attempted to elucidate the complex relationship between happiness (and more generally, affect) and performance in the context of software development. In a study of the affect associated with eliciting requirements, investigating 65 user requirements from two projects, high activation and low pleasure levels were shown to be predictors of high versioning requirements \citep{ColomoPalacios:2011jq}. Pleasure was increased over time with each new version, while activation decreased. 

Theory-building is an important part of software engineering research, and theories regarding affect can inform further empirical studies. One such theory is an explanatory process theory of the impact of affect on development performance~\citep{graziotin2015you}. The theory was formed by a qualitative analysis of the interview data, communications, and observations of two software developers in the same project. The concept of attractors -- affective experiences that earn importance and priority to a developer's cognitive system -- was theorized to have the biggest impact on development performance.

Another theory that relates affective experiences to development performance is the Performance Alignment Work theory~\citep{Fagerholm2015}, which explains a continuous cycle of becoming aware of, interpreting, and adjusting to changing performance demands in and beyond the organizational environment that software development teams are situated in. The theory also explains the ingredients that practitioners consider necessary to form and maintain high-performing teams, which are especially adept at carrying out the cycle. The theory was formed by qualitative analysis of interviews with 16 practitioners in five companies. Affective factors are found throughout the theory, especially in relation to motivators for high performance, and in the social processes of becoming and being part of a high-performing team with a strong identity.

Correlational experiments have found a positive relationship between happiness and positive emotions arising from a development task~\citep{Graziotin2014PEERJ}, and between problem-solving performance and development task productivity~\citep{graziotin2015feelings,Muller2015}. Affect has also been shown to impact debugging performance: in a controlled experiment where participants were asked to write a trace of algorithm execution, induced high pleasure and activation affect were found to be associated with high debugging performance~\citep{Khan2010}.

Further evidence for the link that developers experience between emotion and performance is provided in a survey with 49 developers, assessing emotions they perceived to influence their own productivity~\citep{Wrobel2013}. Positive affective states were perceived to be those that enhance development productivity. The negative affect most prevalently perceived was frustration, which was also the one perceived to deteriorate productivity the most. \citet{Ford2015} have further explored frustration in software engineering through interviews. 67\% of 45 participants in their study reported frustration being a severe issue. Some of the causes for frustration were found to be related to a lack of a good mental model of the code, learning curves of programming tools, too large task sizes, the time required to adjust to new projects, lack of resources, perceived lack of programming experience, inability to complete problems perceived as simple within the bounds of estimated effort, fear of failure, internal hurdles and personal issues, limited time, and issues with peers.

In all the studies investigating the happiness of developers in different forms and its impact on performance, the findings point to a positive relationship. It is similar when software quality is concerned. A series of studies using software repository mining found links between affect, emotions, and politeness, and software quality~\citep{Ortu2015a, Destefanis2016, Mantyla2016a}. Happiness in terms of frequent positive affect and positive emotions was found to be associated with shorter issue fixing time~\citep{Ortu2015a}. The level of arousal, which, when high, is associated with anxiety and burnout, was found to be associated with issue priority~\citep{Mantyla2016a}. Politeness in requests for resolving issues was correlated with shorter resolution time~\citep{Destefanis2016}.

Finally, the present article is part of a series of articles examining a large data set concerning happiness and unhappiness among software developers. We have given a preview of the consequences of unhappiness in a poster paper~\citep{Graziotin:2017} and expanded on the same topic in a workshop paper~\citep{Graziotin:2017semotion}. We found a set of experienced consequences of unhappiness, including detrimental effects on developers' mental well-being, the software development process, and the produced artifacts. The present paper expands on these results. In addition, we have investigated the experienced causes of unhappiness~\citep{Graziotin:2017ease}. We found that the distribution of (un)happiness among software developers, in terms of a quantitative, well established instrument for assessing happiness, pointed towards developers being a slightly happy population, with happiness scores higher than those reported for many other parts of the general human population. Furthermore, we found more than 200 factors representing causes of unhappiness, including being stuck in problem solving, time pressure, bad code quality and coding practice, under-performing colleagues, and feeling inadequate with the task.

In sum, the present state of research shows that happiness and affective experiences play an important role for software developers, and that they are intertwined with software development performance and quality. Several relationships between individual affects and outcomes have been demonstrated. Still, the present research leaves several questions unanswered. On one hand, there is insufficient empirically grounded theory to warrant inquiry into larger sets of causational relationships. On the other hand, the voice of software developers could be expressed more clearly and fully; the understanding of software developers' everyday experiences with the development activity is incomplete. We extend the picture in this paper with further perspectives on happiness and unhappiness among software developers.

\section{Method}
\label{sec:method:method}
The present study is part of a series of inquiries that we conducted on the data from a large-scale survey of developers. The overall research project employs a mixed research method, with elements of both quantitative and qualitative research~\citep{Creswell2009}. The project aims to provide a strong, empirically based assessment of the happiness and unhappiness~\citep{Graziotin:2017ease} of software developers, the causes and consequences of their happiness, and the causes and consequences~\citep{Graziotin:2017semotion,Graziotin:2017} of their unhappiness. That is, all studies share the same general theme and use the same data sample, but have different scopes, variables, and designs. The present study is about the consequences of both happiness and unhappiness, and it is purely qualitative. In this section, we describe both the overall research design as well as methodological details relevant to the present article.

\subsection{Sampling Strategy}
\label{sec:method:data-retrieval}

We consider a software developer to be a person concerned with any aspect of the software construction process (including but not limited to research, analysis, design, programming, testing, and management activities), for any purpose such as work, hobby, or passion. Since we do not accurately know the number of software developers in the world, nor how to reach them, finding a sample that would generalize to the population of software developers is a challenge. We used the GitHub social coding community as an avenue for reaching software developers that would represent the population of developers \textit{well enough}, following several previous studies (e.g., \citet{Gousios:2016hj}). GitHub has more than 30 million visitors each month~\citep{Doll:2015uv} and is, as far as we can tell, the largest social coding community in the world. Software developers using GitHub work on a wide variety of projects, ranging from open source to proprietary software and from solo work to work done in companies and communities.

For the overall research project, we extracted a set of developer contacts from the GitHub Archive \citep{Grigorik:2016}, which stores \emph{public} events occurring in GitHub. We retrieved event data for a period of six months, since we intended to reach developers who were active at the time of our study, meanwhile the amount of events during 6 months is sufficiently large to include as many developers as we desired. We extracted email addresses, given names, company names, developer locations, and the repository name associated with each event. We extracted \emph{unique entries that provided an e-mail address}, resulting in a total of $456\,283$ entries of contact data, including e-mail addresses, given names, company names, and locations of the developers, as well as the repository name related to the public activity.

We discuss the sample size and statistical power for the overall research project in \citep{Graziotin:2017ease}, so we will be brief here. The design of the entire project consists of a quantitative happiness assessment measurement instrument (\citet{Diener2010}; the quantitative results are reported in \citet{Graziotin:2017ease}) and open-ended questions for the causes and consequences of happiness and unhappiness, as well as demographic questions. Given our aim to assess the happiness of the software developers population, the questionnaire item types required us to opt for the most conservative sample size formula by \citet{Yamane:1967tq} for a-priori statistical power calculation (with $\alpha = .01$). Three pilot tests let us estimate a response rate of 2\%, which resulted in a desirable sample of at least $N=664$ complete responses when contacting $33\,200$ randomly selected e-mail addresses. The present paper reports a qualitative study on the consequences of unhappiness and happiness while developing software. Therefore, no statistical power is required.

\subsection{Survey Design}

For the overall research project, we designed a survey consisting of (1) questions regarding demographics, (2) one question with the Scale of Positive and Negative Experience (SPANE, \citet{Diener2010}) with 12 items assessing happiness\footnote{With SPANE, participants report on their affect, expressed with adjectives that individuals recognize as describing emotions or moods, from the past four weeks in order to provide a balance between the sampling adequacy of affect and the accuracy of human memory to recall experiences~\citep{Li2013}, as well as to decrease the ambiguity of people's understanding of the scale itself~\citep{Diener2010}. SPANE is not used in the present study, but we describe it thoroughly in~\citet{Graziotin:2017ease}, including a comparison with competing instruments.}, and (3) two open-ended questions asking for experienced causes and consequences of positive and negative affect when developing software. The present paper deals with the latter, and in particular to only those responses detailing consequences of happiness and unhappiness while developing software.

The two open-ended questions asked developers to recall a time when they were developing software and were experiencing affects described by adjectives in the SPANE instrument (separately for positive and negative affects). Participants were then asked to describe the situation and provide details on what they believed could have caused them to experience the feelings, as well as whether and how their software development was influenced. The two questions appeared on the questionnaire one after the other randomly for mitigating a question-order-effect~\citep{Sigelman:1981fh}. The overall questionnaire is described in an online appendix~\citep{Graziotin:2016ck} with the exception of SPANE, which is freely available at \citep{Diener2009a} but cannot be reproduced here for copyright reasons.

We piloted the questionnaire three times, each with 100 participants from our contact set. The pilots allowed us to estimate and improve response rates by refining the questions and invitation email. No data from the pilots were retained in the final data set and pilot participants did not participate in the final round. The present article covers the results related to the open-ended questions regarding the consequences of positive and negative affect (i.e., happiness and unhappiness) while developing software.

\subsection{Analysis}

Before analysis, we cleaned the data set by removing empty replies, meaningless replies, replies in other languages than English, and replies not relevant to the research questions (e.g. causes for (un)happiness, which are out of scope for this paper). We qualitatively analyzed the cleaned data for the open-ended questions. We developed a coding strategy, applying open coding, axial coding, and selective coding as defined by Corbin and Strauss' Grounded Theory~\citep{Corbin2008}. (For a review and guidelines of Grounded Theory in software engineering research, see \citep{Stol:2016:GTS:2884781.2884833}.) We began by coding consequences of unhappiness. The first three authors each coded the same set of 50 responses using a line-by-line strategy. We then compared the coding structure and strategy and reached an agreement, i.e., a shared axial coding scheme. We took the individual developer as the starting point and unit of observation and analysis, and based the construction of theoretical categories on a model~\citep{Curtis:1988:FSS:50087.50089} of constructs that are \emph{internal} or \emph{external} to the developer. The internal category concerns the developer's own being, while the external category contains artifacts, processes, and people as subcategories. We then divided the data evenly among the first three researchers and proceeded to open code them. We finally merged the codes and conducted a last round of selective coding.

We then used the coding scheme for consequences of unhappiness as a starting point for analyzing the responses related to consequences of happiness, in order to create a comparable scheme. We omitted the detailed leaf categories and created a ``reversed'' or ``mirrored'' set of categories (examples are shown in Table~\ref{tab:category-reversal}). Using this as a starting point, we again divided the data evenly among the first three researchers, and open coded the entire data set, this time looking for consequences of happiness. However, we did not restrict the creation or elimination of categories. Each researcher inserted sub-categories into the coding scheme as required. Thus, if the data would have been significantly different between the the sets of responses, the analysis process would have created two divergent schemes. After the individual open coding phase, we revisited the coding scheme together, and reached an agreement, producing a second shared axial coding scheme for consequences of happiness.

One researcher then selectively re-coded the data that did not fit this coding scheme (only a small number of instances needed re-coding), and the two other researchers verified the coding. Another researcher then made an additional coding pass to add some detail at the leaf level of the coding scheme. Finally, the same researcher revisited both coding schemes and recoded small parts in order to keep categories at similar levels so that the schemes would be comparable in terms of structure and prevalence of in-text occurrences\footnote{We note that this final adjustment means that the coding scheme is slightly different from that reported in the paper we are now extending \citep{Graziotin:2017semotion}; this is intentional and an improvement from the previous paper.}.

The end result is thus two coding schemes that are comparable at the top levels and in terms of structure, while still allowing for differing detailed categories on the more granular level. The coding process also allowed us to keep track of the chain of evidence at all times. An example illustrating the coding process of the consequences of unhappiness is given in Table~\ref{tab:chain-of-evidence}, also available online~\citep{Graziotin:2016ck}.

We note that we only included positive consequences for the happiness part and negative consequences for the unhappiness part. The reverse directions are not explored in this paper, one reason being that they represented less than 5\% of the codes we found. Throughout the process of coding, we monitored our progress and further discussed the coding scheme and strategy in frequent meetings. All qualitative coding and analysis was done using \textit{NVIVO 11}; some calculations were made using \textit{R} and a standard spreadsheet and calculator.

\begin{table*}
\centering
\caption{Examples of reversed categories in coding scheme.}
\label{tab:category-reversal}
\begin{scriptsize}
\begin{tabular}{p{0.45\textwidth} p{0.45\textwidth}}
\toprule
Original category (from consequences of unhappiness) & Reversed category (to consequences of happiness) \\
\midrule
developer's own being$\,\to\,$low cognitive performance & developer's own being$\,\to\,$high cognitive performance \\
developer's own being$\,\to\,$low motivation & developer's own being$\,\to\,$high motivation \\
developer's own being$\,\to\,$work withdrawal & developer's own being$\,\to\,$high work engagement and perseverance \\
external consequences$\,\to\,$artifact and working with artifact$\,\to\,$low code quality & external consequences$\,\to\,$artifact and working with artifact$\,\to\,$high code quality \\
external consequences$\,\to\,$process$\,\to\,$low productivity & external consequences$\,\to\,$process$\,\to\,$high productivity \\
external consequences$\,\to\,$process$\,\to\,$decreased process adherence & external consequences$\,\to\,$process$\,\to\,$increased process adherence \\
\bottomrule
\end{tabular}
\end{scriptsize}
\end{table*}

\begin{table*}
\centering
\caption{Example of coding phases showing the chain of evidence. The first column contains a response fragment. The next three columns show how three different researchers open coded the fragment. The fourth column shows how the open codes were merged during the agreement phase. The remaining columns show the higher-level codes of the axial and selective coding phases.}
\label{tab:chain-of-evidence}
\begin{scriptsize}
\begin{tabular}{>{\raggedright}p{2.5cm} >{\raggedright}p{2.1cm} >{\raggedright}p{2.1cm} >{\raggedright}p{2.1cm} >{\raggedright}p{2.7cm} >{\raggedright}p{2cm} >{\raggedright\arraybackslash}p{2.1cm}}
\toprule
Response fragment & Open coding R1 & Open Coding R2 & Open Coding R3 & Open Coding Merged & Axial Coding & Selective Coding \\
\midrule
I am annoyed when a coworkers lack of programming progress limits my programming progress. &
\mbox{} \newline Team member not progressing &
\mbox{} \newline Waiting for others to finish their code &
Coding \newline \newline Collaboration &
Coordination issue \newline \newline Bad code written by others &
Process \newline \newline Code and coding &
Process \newline \newline Artifact and working with artifact
\\
\midrule
I am angry when when I work on code for a long time and it does not work. [\ldots] Eventually I have to walk away to reflect. &
\mbox{} \newline Staying on current broken code &
\mbox{} \newline Working for a long time on code that does not work &
\mbox{} \newline \newline \newline \newline \newline \newline Negative outcome &
Unexplained broken code &
Code and coding \newline \newline \newline \newline \newline Work withdrawal &
Artifact and working with artifact \newline \newline \newline Individual (Consequences)
\\
\midrule
I am anxious when I am put in charge of a broken product and feel that massive restructuring is required. I have to justify the major destruction of 'working' code. &
Anxiety \newline \newline Handling a broken project \newline \newline Feeling that code must be restructured &
\mbox{} \newline \newline \newline \newline \newline \newline \newline Having to justify code restructuring &
\mbox \newline \newline \newline Working with legacy &
Anxiety \newline \newline Bad code quality and coding practices &
Mental unease or disorder \newline \newline Code and coding &
Individual (Consequences) \newline \newline Artifact and working with artifact
\\
\bottomrule
\end{tabular}
\end{scriptsize}
\end{table*}

\section{Results}
\label{sec:results:results}

In this section, we summarize the results of our investigation. We first show descriptive statistics describing the demographics of the participants. We then proceed to the qualitative data related to our RQs. We summarize the elicited consequences of unhappiness while developing software first, then move to the consequences of happiness, reflecting the order in which the analysis was performed.

\subsection{Descriptive Statistics}
\label{ssec:results:descriptive}

A total of $2\,220$ individuals participated in the overall quantitative and qualitative study (see \citet{Graziotin:2017ease}). Out of these, $1\,318$ provided data for the open questions on causes and consequences of happiness and unhappiness, of which $317$ provided answers pertaining to the present study of consequences of happiness or consequences of unhappiness.

Since we treated the response data for the two research questions separately, the set of respondents for each question is different. In total, there are $317$ respondents in the two data sets combined, counting each respondent only once. $70$ of the respondents (22\%) occur in both data sets, while $247$ respondents (78\%) occur only in one of the data sets. The descriptive statistics for the the data is shown in Table~\ref{tab:descriptive-statistics} and discussed below.

We obtained $181$ valid and complete responses related to RQ1, which resulted in $172$ male participants (95\%) and 8 female (4\%). The remaining participant indicated \textit{other / prefer not to disclose} for gender. The mean year of birth was 1984 (standard deviation, $sd=8.27$), while the median was 1986. A wide range of nationalities was represented, comprising 45 countries. 141 (78\%) participants were professional software developers, 7\% were students, and 13\% were in other roles (such as manager, CEO, CTO, and academic researcher). The remaining participants were unemployed and not students. The participants declared a mean of 8.22 years ($sd=7.83$) of software development working experience; the median was 5 years.

We obtained $206$ valid and complete responses related to RQ2. In that set, $195$ participants were male (95\%) and 9 were female (4\%). The remaining two participants declared their gender as \textit{other / prefer not to disclose}. The mean year of birth was 1985 ($sd=9.67$), and the median was 1987. A wide range of nationalities was represented, with 53 countries. 153 (74\%) participants were professional software developers, 9\% were students, and 16\% had other roles (defined as above). The remaining two participants (1\%) were unemployed and not students. The participants reported a mean of 8.23 years ($sd=8.9$) of software development working experience, with a median of 5 years.

\newlength{\mytabcolsep}
\setlength{\mytabcolsep}{\tabcolsep}
\setlength{\tabcolsep}{2.5pt}
\begin{table*}
\centering
\caption{Descriptive statistics for the data obtained for each research question. N denotes the size of each data set. Percentages are relative to the size of the data for each question. The data sets partly overlap; 70 data points (22\%) are in both data sets.}
\label{tab:descriptive-statistics}
\begin{scriptsize}
\begin{tabular}{c r r r r r r r r r r r r r r r}
\toprule
\mbox{} &
\mbox{} &
\multicolumn{3}{c}{Gender} &
\multicolumn{3}{c}{Year of birth} &
\mbox{} &
\multicolumn{4}{c}{Role} &
\multicolumn{3}{c}{Work experience (years)} \\

RQ &
\multicolumn{1}{c}{N} &
\multicolumn{1}{c}{Male} &
\multicolumn{1}{c}{Female} &
\multicolumn{1}{c}{Other} &
\multicolumn{1}{c}{Mean} &
\multicolumn{1}{c}{Median} &
\multicolumn{1}{c}{Std.\ dev.} &
\multicolumn{1}{c}{Countries} &
\multicolumn{1}{c}{Professional} &
\multicolumn{1}{c}{Student} &
\multicolumn{1}{c}{Unemployed} &
\multicolumn{1}{c}{Other} &
\multicolumn{1}{c}{Mean} &
\multicolumn{1}{c}{Median} &
\multicolumn{1}{c}{Std.\ dev.} \\
\midrule

1 & 181 & 172 (95\%) & 8 (4\%) & 1 (1\%) & 1984 & 1986 & 8.27 & 45 & 141 (78\%) & 12 (7\%) & 5 (3\%) & 23 (13\%) & 8.22 & 5 & 7.83 \\
2 & 206 & 195 (95\%) & 9 (4\%) & 2 (1\%) & 1985 & 1987 & 9.67 & 53 & 153 (74\%) & 18 (9\%) & 2 (1\%) & 33 (16\%) & 8.23 & 5 & 8.9 \\

\bottomrule
\end{tabular}
\end{scriptsize}
\end{table*}
\setlength{\tabcolsep}{\mytabcolsep}

\subsection*{What are the Experienced Consequences of Unhappiness Among Software Developers While Developing Software?}
\label{ssec:resultsrq2}

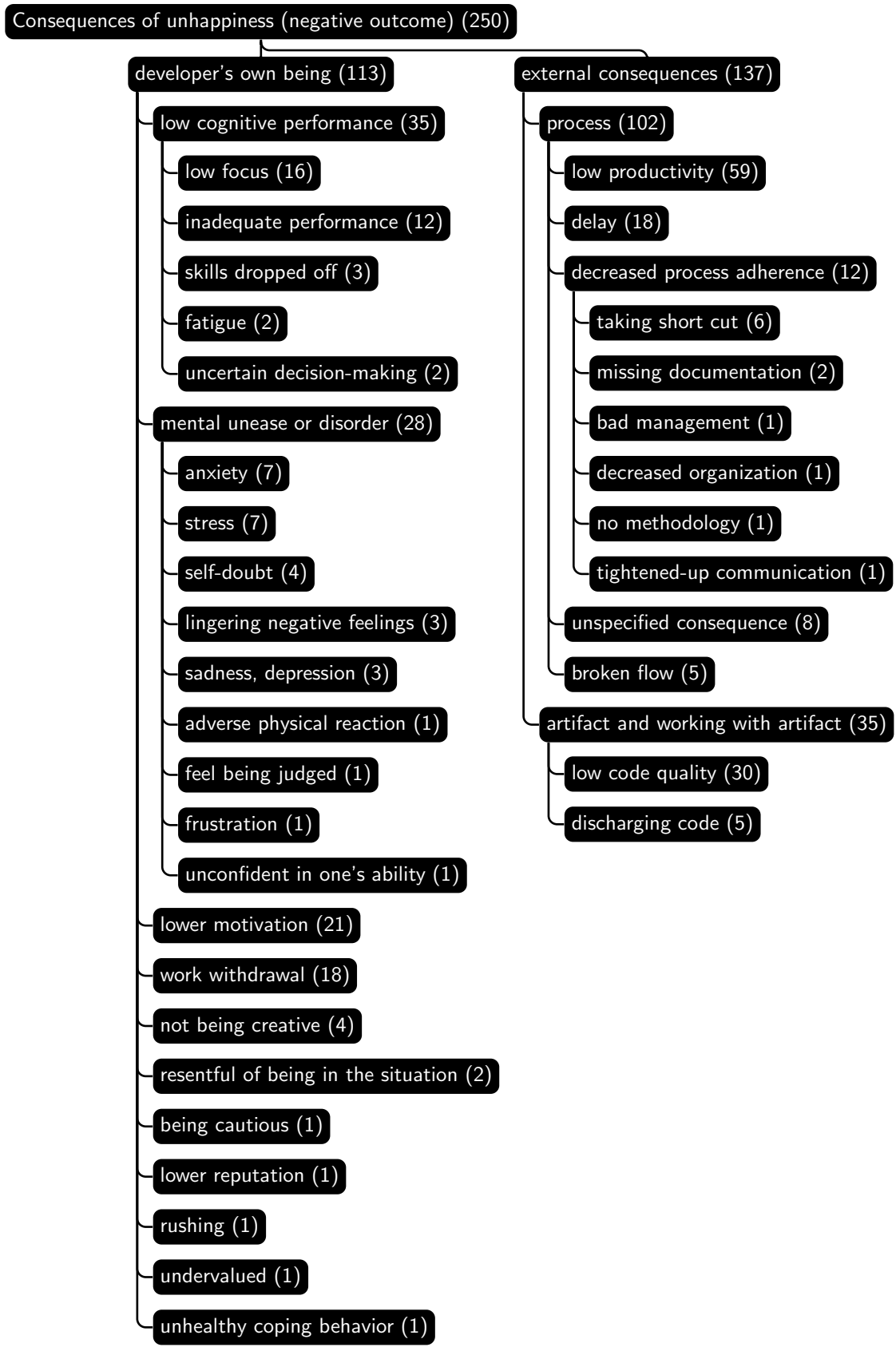
\begin{figure*}
\centering
\begin{forest}
my label/.style={
    label={[font=\normalsize\sffamily]right:{#1}},
},
where level=1 {
    child anchor=north,
    !u.parent anchor=south,
    before computing xy={
        l*=.5,
    },
    if={n==(int((n_children("!u")+1)/2))}{
        calign with current edge
    }{},
    edge path'={(!u.parent anchor) -- ++(0, -5pt) -| (.child anchor)},
    for tree={
        folder,
        font=\normalsize\sffamily,
        text=black,
        if level=0
            {fill=cyan!80}
            {fill/.wrap pgfmath arg={cyan!#1}{int(80-10*(mod((level()-1),4)))}},
        rounded corners=4pt,
        grow'=0,
        edge={black, rounded corners, line width=1pt},
        fit=band,
    },
}{},
for tree={
    font=\normalsize\sffamily,
    text=white,
    fill=black,
    rounded corners=4pt,
    edge={black, rounded corners, line width=1pt},
},
[Consequences of unhappiness (negative outcome) (250)
    [developer's own being (113)
        [low cognitive performance (35)
            [low focus (16)]
            [inadequate performance (12)]
            [skills dropped off (3)]
            [fatigue (2)]    
            [uncertain decision-making (2)]
        ]        
        [mental unease or disorder (28)
	        [anxiety (7)]
	        [stress (7)]
	        [self-doubt (4)]
	        [lingering negative feelings (3)]
	        [sadness{,} depression (3)]
	        [adverse physical reaction (1)]
	        [feel being judged (1)]
	        [frustration (1)]
	        [unconfident in one's ability (1)]
        ]      
        [lower motivation (21)]
        [work withdrawal (18)]
        [not being creative (4)]
        [resentful of being in the situation (2)]
        [being cautious (1)]        
        [lower reputation (1)]                
        [rushing (1)]
        [undervalued (1)]
        [unhealthy coping behavior (1)]
    ]
    [external consequences (137)
        [process (102)
			[low productivity (59)]
			[delay (18)]            
			[decreased process adherence (12)
				[taking short cut (6)]
				[missing documentation (2)]
				[bad management (1)]
				[decreased organization (1)]
				[no methodology (1)]
				[tightened-up communication (1)]
			]
			[unspecified consequence (8)]
			[broken flow (5)]
		]
        [artifact and working with artifact (35)
			[low code quality (30)]
            [discharging code (5)]
        ]
    ]
]
\end{forest}
\caption{Categories for consequences of unhappiness. The numbers indicate the amount of coded instances at each level, including sub-categories.}
\label{fig:categories-unhappiness}
\end{figure*}

We identified 250 coded instances related to the negative consequences of unhappiness. They were grouped into a scheme with 42 categories. The structure of the scheme is shown in Figure~\ref{fig:categories-unhappiness}, which is also available as archived open data~\citep{Graziotin:2016ck}. The material was divided into an internal category, labeled \emph{developer's own being} (113 references), and the external categories \emph{process} (102) and \emph{artifact} (35). The top ten categories (with tied categories counted as one) are shown in Table~\ref{tbl:top10unhappiness}.

\begin{table*}
\caption{Top 10 Consequences of Unhappiness, Categories, and Frequency. Tied consequences are counted as one.}
\label{tbl:top10unhappiness}
\centering
\begin{small}
\begin{tabular}{llr}
\toprule
Consequence & Category &  Freq. \\
\midrule
Low productivity & external consequences$\,\to\,$process & 59 \\
Low code quality & external consequences$\,\to\,$artifact and working with artifact & 30 \\
Lower motivation & developer's own being & 21 \\
Work withdrawal & developer's own being & 18 \\
Delay & external consequences$\,\to\,$process & 18 \\
Low focus & developer's own being$\,\to\,$low cognitive performance & 16 \\
Inadequate performance & developer's own being$\,\to\,$low cognitive performance & 12 \\
Decreased process adherence & external consequences$\,\to\,$process & 12 \\
Unspecified consequence & external consequences$\,\to\,$process & 8 \\
Anxiety & developer's own being$\,\to\,$mental unease or disorder & 7 \\
Stress & developer's own being$\,\to\,$mental unease or disorder & 7 \\
Discharging code & external consequences$\,\to\,$artifact and working with artifact & 5 \\
Broken flow & external consequences$\,\to\,$process & 5 \\
Self-doubt & developer's own being$\,\to\,$mental unease or disorder & 4 \\
Not being creative & developer's own being & 4 \\
\bottomrule
\end{tabular}
\end{small}
\end{table*}

\subsection{Internal Consequences---Developer's Own Being}
\label{sssec:resultsrq2:individual_consequences}

We refer to the internal consequences as pertaining to the \textit{developers's own being}, as we want to highlight that these consequences are directed towards and alter the self rather than merely being situated within the person. The factors related to the developer's own being do not demonstrate a clear structure. This to some extent reflects the versatile states of mind of developers and the feelings they could have while they develop software.

The most significant consequences of unhappiness for the \textit{developers' own being} are, in terms of frequency: low cognitive performance, mental unease or disorder, low motivation and work withdrawal.

\textbf{Low cognitive performance} is a category to group all those consequences related to low or inadequate mental performance. Specific symptoms include low focus: ``\textit{[\ldots] the negative feelings lead to not thinking things through as clearly as I would have if the feeling of frustration was not present}''; cognitive skills dropping off: ``\textit{My software dev skills dropped off as I became more and more frustrated until I eventually closed it off and came back the next day to work on it}''; and general mental fatigue: ``\textit{Getting frustrated and sloppy}''.

The \textbf{mental unease or disorder} category collects all those consequences that threaten mental health\footnote{\label{trained-psych}In this study, we report what the participants stated, but we remind readers that only trained psychologists and psychiatrists should treat or diagnose mental disorders.}. Apart from general lingering negative feelings, the participants reported that unhappiness while developing software is a cause of, in order of frequency, anxiety: ``\textit{These kinds of situations make me feel panicky}''; stress: ``\textit{[\ldots] only reason of my failure due of burnout}''; self-doubt: ``\textit{If I feel particularly lost on a certain task, I may sometimes begin to question my overall ability to be a good programmer}''; and sadness and depression. Participants mentioned depression as feeling depressed, e.g., ``\textit{feels like a black fog of depression surrounds you and the project}'' or ``\textit{I get depressed}''. In addition, feelings of being judged, frustration, lack of confidence in one's ability (overlapping with, but slightly different from, self-doubt; a generalised self-efficacy belief), and (even) adverse physical reaction, although mentioned only once as the perceived consequences of unhappiness, do appear, and manifest the extent of mental unease or disorder caused by unhappy feelings.

\textbf{Low motivation} is also an important consequence of unhappiness for software developers. Motivation is a set of psychological processes that cause the mental activation, direction, and persistence of voluntary actions that are goal directed \citep{Mitchell1982}. Motivation has been the subject of study in software engineering literature (e.g, \citet{Franca2014b}), and we reported that affective experiences are related to motivation even though they are not the same construct \citep{graziotin2015affect}. The participants were clear in stating that unhappiness leads to low motivation for developing software, e.g., ``\textit{[the unhappiness] has left me feeling very stupid and as a result I have no leadership skills, no desire to participate and feel like I'm being forced to code to live as a kind of punishment. [\ldots]}'', or ``\textit{Also, I'm working at a really slow pace [\ldots] because I'm just not as engaged with the work}''.

\textbf{Work withdrawal} is a very destructive consequence of unhappiness, and it emerged often among the responses. Work withdrawal is a family of behaviors that is defined as employees' attempts to  remove themselves, either temporarily or permanently, from quotidian work tasks \citep{Miner2010}. The gravity of this consequence ranged from switching to another task, e.g., ``\textit{[\ldots] you spend like 2 hours investigating on Google for a similar issue and how it was resolved, you find nothing, desperation kicks in. It clouds your mind and need to do other things to clear it}'', to considering quitting developing software, ``\textit{I really start to doubt myself and question whether I'm fit to be a software developer in the first place}'', or even, ``\textit{I left the company}''.

Other consequences perceived by the participants include not being creative, ``\textit{it is very difficult for me to do any creative work when angry or unhappy}'', resentful of being in the situation, being cautious, lower reputation, rushing, undervalued, and unhealthy coping behavior, such as smoking excessively. Because of the low frequency of mentions in the data, we do not elaborate on these consequences further here.

\subsection{External Consequences---Process }

The category of \textit{process} collects those unhappiness consequences that are related to a software development process, endeavour, or set of practices that is not explicitly tied up to an artifact (see Section \ref{sssec:resultsrq2:artifactoriented_consequences}).

\textbf{Low productivity} is a category for grouping all consequences of unhappiness related to performance and productivity losses\footnote{See \citep{graziotin2015you} and \citep{Fagerholm2015} for our stance on a definition of productivity and performance in software engineering.}. The codes within this category were ranging from very simple and clear ``\textit{productivity drops}'', ``\textit{[negative experience] definitely makes me work slower}'' to more articulated ``\textit{[unhappiness] made it harder or impossible to come up with solutions or with good solutions}'', ``\textit{[\ldots], and [the negative experience] slowed my progress because of the negative feeling toward the feature}''.

Unhappiness was reported to be causing \textbf{delay} in executing process activities: ``\textit{In both cases [negative experiences] the emotional toll on me caused delays to the project}''. Unhappiness causes glitches to communication activities and a disorganized process: ``\textit{Miscommunication and disorganization made it very difficult to meet deadlines}''.

Developers declared that unhappiness caused them to have \textbf{decreased process adherence}, i.e., deviating from the agreed set of practices. Specifically, unhappiness was reported to lead developers to compromise in terms of actions, in order to just get rid of the job: ``\textit{In these instances my development tended towards immediate and quick `ugly' solutions}''.
Developers see the quality of their code compromised (Section \ref{sssec:resultsrq2:artifactoriented_consequences}) but also decide to take shortcuts when enacting a software process, compromising methodology, good management, and the quality of the process itself: ``\textit{[\ldots] can lead to working long hours and trying to find shortcuts. I'm sure this does not lead to the best solution, just a quick one}''. The process adherence can suffer due to communication aspects, too: ``\textit{my development was influenced by [negative affect] in that it caused me to tighten up communications and attempt to force resolution of the difficulties}''.

The \textbf{broken flow} category is related to the process deviation in terms of process unevenness and wasted time in restarting tasks. The concept of flow has been defined by \citet{Csikszentmihalyi1997} as a state of intense attention and concentration resulting from task-related skill and challenge being in balance. Flow has been investigated by \citet{Muller2015} in the context of software development. Unhappiness causes interruptions in developers' flow, resulting in adverse effects on the process. As put by a participant, `\textit{things like that [of unhappiness] often cause long delays, or cause one getting out of the flow, making it difficult to pick up the work again where one has left off. }''. Unhappiness and broken flow make developers stand up and ``\textit{[\ldots] make me quit and take a break}''; the feeling of getting stuck is persistent.

\subsection{External Consequences---Artifact-oriented}
\label{sssec:resultsrq2:artifactoriented_consequences}
The category of \textit{artifact-oriented} consequences groups all those consequences that are directly related to a development product, e.g., software code, requirements, and to working with it. As expected by the foci of previous research, the most important consequence of unhappiness of software developers was low software quality.

\textbf{Low code quality} represents the consequences of unhappiness of developers that are related to deterioration of the artifacts' quality. The participants reported that ``\textit{eventually [due to negative experiences], code quality cannot be assured. So this will make my code messy and more bug can be found in it}'', but also mentioned making the code less performant, or ``\textit{As a result my code becomes sloppier}''. Moreover, participants also felt that they could discharge quality practices, e.g., ``\textit{[\ldots] so I cannot follow the standard design pattern}'', as a way to cope with the negative experiences.

\textbf{Discharging code} could be seen as an extreme case of productivity and quality drop. Participants were very clear that they were not referring to some refactoring strategies but really meant relieving themselves from a charge, load, or burden (c.f.~\citet{merriam2017}). We found some instances of participants who destroyed the task-related codebase, e.g., ``\textit{I deleted the code that I was writing because I was a bit angry}'', up to deleting entire projects: ``\textit{I have deleted entire projects to start over with code that didn't seem to be going in a wrong direction}''.

\subsection*{What are the Experienced Consequences of Happiness Among Software Developers While Developing Software?}
\label{ssec:resultsrq1}

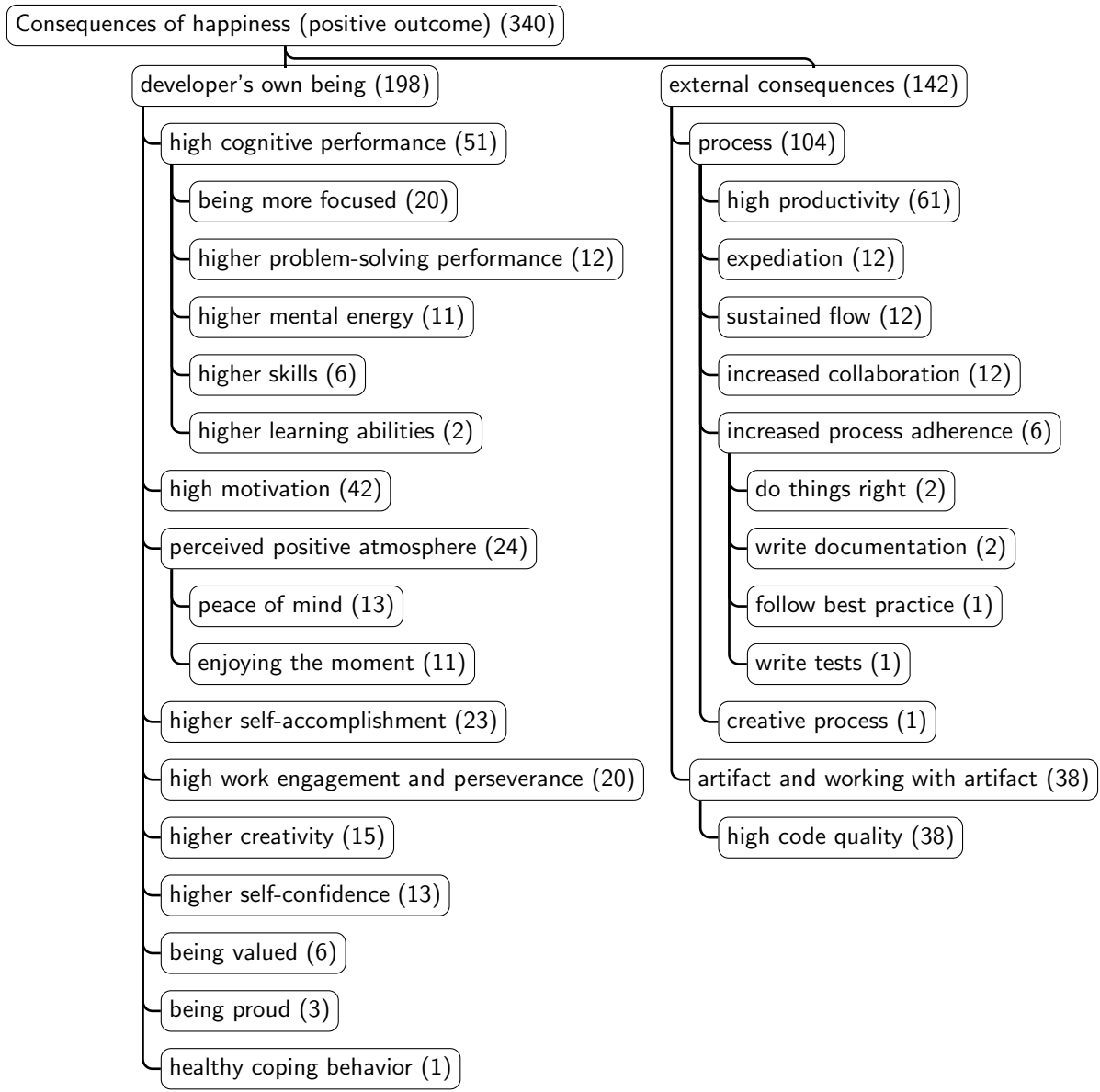
\begin{figure*}
\centering
\begin{forest}
my label/.style={
    label={[font=\normalsize\sffamily]right:{#1}},
},
where level=1 {
    child anchor=north,
    !u.parent anchor=south,
    before computing xy={
        l*=.5,
    },
    if={n==(int((n_children("!u")+1)/2))}{
        calign with current edge
    }{},
    edge path'={(!u.parent anchor) -- ++(0, -5pt) -| (.child anchor)},
    for tree={
        folder,
        font=\normalsize\sffamily,
        text=black,
        if level=0
            {fill=cyan!80}
            {fill/.wrap pgfmath arg={cyan!#1}{int(80-10*(mod((level()-1),4)))}},
        rounded corners=4pt,
        grow'=0,
        edge={black, rounded corners, line width=1pt},
        fit=band,
    },
}{},
for tree={
    font=\normalsize\sffamily,
    text=black,
    fill=white,
    rounded corners=4pt,
    edge={black, rounded corners, line width=1pt},
    draw={black, thin},
},
[Consequences of happiness (positive outcome) (340)
    [developer's own being (198)
        [high cognitive performance (51)
            [being more focused (20)]
            [higher problem-solving performance (12)]
            [higher mental energy (11)]
            [higher skills (6)]
            [higher learning abilities (2)]
        ]
        [high motivation (42)]
        [perceived positive atmosphere (24)
            [peace of mind (13)]
            [enjoying the moment (11)]
        ]
        [higher self-accomplishment (23)]
        [high work engagement and perseverance (20)]
        [higher creativity (15)]
        [higher self-confidence (13)]
        [being valued (6)]
        [being proud (3)]
        [healthy coping behavior (1)]
    ]
    [external consequences (142)
        [process (104)
            [high productivity (61)]
            [expediation (12)]
            [sustained flow (12)]
            [increased collaboration (12)]
            [increased process adherence (6)
	            [do things right (2)]
	            [write documentation (2)]
	            [follow best practice (1)]
	            [write tests (1)]           
            ]
            [creative process (1)]
        ]
        [artifact and working with artifact (38)
            [high code quality (38)]
        ]
    ]
]
\end{forest}
\caption{Categories for consequences of happiness. The numbers indicate the amount of coded instances at each level, including sub-categories.}
\label{fig:categories-happiness}
\end{figure*}

We now provide a summary of the elicited positive consequences of happiness while developing software. We identified a total of 340 coded instances related to the consequences of happiness, They are grouped into 32 categories and sub-categories. The structure of the categories is shown in Figure~\ref{fig:categories-happiness}, and is available as archived open data~\citep{Graziotin:2016ck}. Consistent with the unhappiness consequences, we defined an internal category, \emph{developer's own being} (198 coded instances), and two external categories, \emph{process} (104) and \emph{artifact} (38). The top ten categories (with tied categories counted as one) are shown in Table~\ref{tbl:top10happiness}.

\begin{table*}
\caption{Top 10 Consequences of Happiness, Categories, and Frequency. Tied consequences are counted as one.}
\label{tbl:top10happiness}
\centering
\begin{small}
\begin{tabular}{llr}
\toprule
Consequence & Category &  Freq. \\
\midrule
High productivity & external consequences$\,\to\,$process & 61 \\
High motivation & developer's own being & 42 \\
High code quality & external consequences$\,\to\,$artifact and working with artifact & 38 \\
Higher self-accomplishment & developer's own being & 23 \\
High work engagement and perseverance & developer's own being & 20 \\
Being more focused & developer's own being$\,\to\,$high cognitive performance & 20 \\
Higher creativity & developer's own being & 15 \\
Higher self-confidence & developer's own being & 13 \\
Peace of mind & developer's own being$\,\to\,$perceived positive atmosphere & 13 \\
Increased collaboration & external consequences$\,\to\,$process & 12 \\
Sustained flow & external consequences$\,\to\,$process & 12 \\
Expediation & external consequences$\,\to\,$process & 12 \\
Higher problem-solving performance & developer's own being$\,\to\,$high cognitive performance & 12 \\
Enjoying the moment & developer's own being$\,\to\,$perceived positive atmosphere & 11 \\
Higher mental energy & developer's own being$\,\to\,$high cognitive performance & 11 \\
Increased process adherence & external consequences$\,\to\,$process & 6 \\
Being valued & developer's own being & 6 \\
Higher skills & developer's own being$\,\to\,$high cognitive performance & 6 \\
\bottomrule
\end{tabular}
\end{small}
\end{table*}

\subsection{Internal Consequences---Developer's Own Being}
\label{sssec:resultsrq1:internal_consequences}

We now turn to the internal consequences of happiness, which we call \textit{developer's own being} for the same reason we stated at the beginning of Section \ref{sssec:resultsrq2:individual_consequences}.

Similar to those of unhappiness consequences, the set of factors related to the developer's own being is varied and displays no clear structure. The most significant consequences, in terms of frequency, of happiness for the \textit{developer's own being} are: high cognitive performance, high motivation, perceived positive atmosphere, higher self-accomplishment, high work engagement and perseverance, higher creativity and higher self-confidence. Other less mentioned consequences include being valued, being proud, and healthy coping behavior. 

\textbf{High cognitive performance} is a category grouping all consequences related to high mental performance, such as being focused: ``\textit{My software development is influenced because I can be more focused on my tasks and trying to solve one problem over another}''; higher problem-solving performance: ``\textit{I mean, I can write codes and analyze problems quickly and with lesser or no unnecessary errors when I'm not thinking of any negative thoughts}''; higher mental energy: ``\textit{This influenced my work by making me more alert, concentrated}''; higher skills: ``\textit{\ldots felt that my skill has improved tremendously}''; and higher learning abilities: ``\textit{It made me want to pursue a masters in Computer science and learn interesting and clever ideas to solve problems}''. We excluded creativity from this category; while cognitive performance is an important component of creativity, the latter also requires divergent thinking and emotion plays a different role than in more logical tasks (see, e.g., \citet{baas2008} and \citet{davis2009} for discussions on the emotional aspects of creativity).

\textbf{High motivation} is an important consequence of happiness for software developers. The participants stated that increased motivation occurred as they were happy, e.g., ``\textit{When I write a bunch of code and compile/run it and it works without any errors. Though, when that happens I also become a bit suspicious. I felt more motivated to continue writing code at that point}''.

\textbf{Perceived positive atmosphere} is an internal evaluation of the surrounding social context that participants reported occurring with happiness. Participants reported greater peace of mind: ``\textit{it's very comforting}''; as well as simply enjoying the moment: ``\textit{\ldots gives you the sense of achievement and joy}''.

\textbf{Higher self-accomplishment} refers to stronger or more frequent feelings of having achieved something successfully. With increased happiness during development, respondents reported that they felt having performed something successfully: ``\textit{The sense of accomplishment when finishing something that actually works is very rewarding}''.

\textbf{High work engagement and perseverance} was reported to occur when respondents were happy. This means, e.g., pushing forward with tasks: ``\textit{I think I was more motivated to work harder the next few hours}''.

\textbf{Higher creativity} was also a reported result of happiness. Participants reported finding it easier to come up with new ideas and to think divergently: ``\textit{This give you energy [which] feed your creativity and you come up [with] more crazy and wonderful ideas}''.

\textbf{Higher self-confidence} means greater trust in one's personal abilities. This category includes both higher general self-confidence as well as task-specific self-confidence (self-efficacy). Participants expressed a shift in self-confi\-dence when they were happy, e.g.: ``\textit{These situations encourage me to pick up tasks that I was afraid of before, because they seemed to be too difficult}''.

\textbf{Being valued} is another social evaluation experienced by participants when being happy: ``\textit{My boss said `You're the best' and then gave me a hug and I said `nuh uh your tha best'}''.

Participants also reported \textbf{being proud} when they were happy, both directed towards their work -- ``\textit{great pride in the work I've just completed}'' -- and directed towards themselves -- ``\textit{I get more proud of myself}''.

Finally, one participant reported on another developer's happiness encouraging \textbf{healthy coping behavior}: ``\textit{[He] has a definite positive effect on development, not just the production side but also my attitude when dealing with negative issues}''.

\subsection{External Consequences---Process}
\label{sssec:resultsrq1:external_consequences:process}

Being happy was often associated by the participants to several positive consequences closely related to software development processes. Among different consequences, \textbf{high productivity} is the most frequently listed one, followed by \textbf{expediation}, \textbf{sustained flow}, \textbf{increased collaboration} and \textbf{increased process adherence}. \textbf{Creative process} is also a perceived consequence of happiness, though with one occurrence only.

The most noticeable positive consequence related to process by far is \textbf{high productivity}, as put by several participants, ``\textit{When I have this [happy] feeling I can just code for hours and hours}'', ``\textit{I felt that my productivity grew while I was happy}'', ``\textit{The better my mood, the more productive I am}''. One participant described in more detail such high productivity caused by happiness: ``\textit{I become productive, focused and enjoy what I'm doing without wasting hours looking here and there in the code to know how things are hooked up together}''. One interesting aspect of this high productivity caused by being in a happy state is that the developers tend to take on undesired tasks. As one participant admitted: ``\textit{I think that when I'm in this happy state I am more productive. The happier I am the more likely I'll be able to accomplish tasks that I've been avoiding}''. Another intriguing aspect is the long-term consideration invoked by the happy state of mind: ``\textit{I find that when I feel this way [being happy], I'm actually more productive going into the next task and I make better choices in general for the maintenance of the code long-term. [\ldots] I'm more likely to comment code thoroughly}''.

\textbf{Expediation} as a perceived consequence emphasizes that when developers feel good during development, tasks can be sped up without sacrificing quality, and ``\textit{it seems more likely to reach my goals faster}''. Meantime, developers can enter a state of a \textbf{sustained flow}. Developers feel being in a state of flow, full of energy and with strong focus. In such a state, they are ``\textit{unaware of time passing}''. They can ``\textit{continue to code without anymore errors for the rest of the day}'', and ``\textit{just knock out lines of code all day}'', with ``\textit{dancing fingers, my code is like a rainbow}''. 

Happy developers can also mean more collaborative team members, leading to \textbf{increased collaboration}. This is reflected under several different aspects. We saw quite a repeating pattern that happiness leads to more willingness to share knowledge (\textit{``I'm very curious and i like to teach people what i learned''}). Being happy leads to willingness to join peers in solving a problem (\textit{``we never hold back on putting our brains together to tackle a difficult problem or plan a new feature''}) even when not related to the task at hand or the current responsibilities (\textit{``I was more willing to help them with a problem they were having at work.''}). Additionally, participants stressed out how \textit{``the interactions with co-workers and office-mates was almost always enjoyable''} and that happiness leads to \textit{``good manners, practice and planning''}.

Positive feelings could influence the practices that a team uses and promote good manners where team members work with each other, in turn making the interactions among team members enjoyable.

Being more self-disciplined, as indicated by \textbf{increased process adherence}, is also perceived by the participants as a consequence of being happy, as demonstrated by this response: ``\textit{when I am happy to work, I usually try new things and follow best practices and standards as much as possible}''. The adherence to process is manifested especially when testing and documentation are concerned: ``\textit{the better I feel, I'm more likely to produce elegant code, with tests and documentation}''.

\textbf{Creative process} can also be a positive consequence of developers being happy, as stated in this response: ``\textit{if [\ldots] I have a general good mood, the software process gets to be creative and very good}''.

\subsection{External Consequences---Artifact-oriented}
\label{sssec:resultsrq1:external_consequences:artefact}

\textbf{High code quality} is the single most significant consequence, in terms of frequency, of happy feelings while developing software, as perceived by the participants. A participant told a small story about their work: ``\textit{I was building an interface to make two applications talk. It was an exciting challenge and my happy and positive feelings made me go above and beyond to not only make it functional but I made the UX nice too. I wanted the whole package to look polished and not just functional}''. Higher quality of code is generally realized when developers are happy, because they tend to make less mistakes, see solutions to problems more easily, and make new connections to improve the quality of the code. A participant argued: ``\textit{When I'm in a good mood and I feel somehow positive, the codes I write seems to be very neat and clean. I mean, I can write codes and analyze problems quickly and with lesser or no unnecessary errors}''. As a result, the code is cleaner, more readable, better commented and tested, and with less errors and bugs.

\section{Discussion}\label{sec:discussion:discussion}

The main finding of our study is that software developers experience several consequences of unhappiness and happiness, with most of the consequences of unhappiness being external (55\%) and most consequences of happiness pertaining to the developer's own being (58\%) (internal consequences). We summarise the ten most frequent categories on the leaf level in tables~\ref{tbl:top10unhappiness} and~\ref{tbl:top10happiness}. We may say that broadly speaking, developers more frequently consider their happiness to benefit themselves, and their unhappiness to be detrimental to others. However, this applies on the most general level of our results, and examining the details reveals several important findings. In this section, we first consider the answers to our research questions, examine the implications of the research questions taken together, discuss the limitations of the study, and provide recommendations for practitioners and researchers.

\subsection{Addressing the research questions}

\paragraph{Consequences of Unhappiness}

Our analysis to answer RQ1 resulted in 42 categories of negative consequences that developers experienced to stem from unhappiness. Many of these have a detrimental impact on several important software engineering outcomes. Productivity and performance are the aspects which suffer most from unhappy developers. When grouping codes for low cognitive performance and process-related productivity, approximately 55\% of the related in-text references deal with productivity and performance drops. Those results are in line with and support the related work in software engineering research \citep{graziotin2015you, Graziotin2014PEERJ, graziotin2015feelings, Muller2015, Khan2010, Wrobel2013} which quantified the relationship or attempted to explain the link.

Slightly more than half of the in-text occurrences (55\%) concern external consequences. Most of these relate to the process of developing software. Unhappiness is reported to result in unevenness in the process: low productivity, delays, and broken flow. Unhappiness decreases developers' adherence to the agreed process, resulting in taking process-related shortcuts (i.e., to ``cut corners''). These deviations are often mentioned to cause issues in terms of software quality. External consequences were also reported on software artifacts, where, in addition to low code quality, discharging code was mentioned. While a few studies have been conducted on the impact of developers' affect on software quality (e.g, \citet{Khan2010,Destefanis2016,Ortu2015a,Ortu:2016gz}), we encourage further research on the matter.

Slightly less than half of the in-text occurrences (45\%) concern consequences for the developer's own being (internal consequences). The most prevalent consequences in this category are related to the cognitive performance of developers. Unhappiness takes its toll in terms of low focus, inadequate performance, reduction of skills, fatigue, and problems with decision-making. Such consequences hit directly at the core of the software development activity, as it is inherently intellectual. The link to the external consequences related to low productivity, decreased process adherence, broken flow, and low quality appears obvious, but would have to be confirmed in further studies.

Our results further show that unhappiness while performing software development may be a source of several mental-related issues that are known to be of detrimental effect to the individual and the work environment. We found situations of mental unease, e.g., low self-esteem, high anxiety, burnout, and stress. Initial software engineering research on the latter two has started (e.g., \citet{Mantyla2016a}), but the related work in psychology is comprehensive and alarming in regards to how disruptive these issues are on well-being. Furthermore, our data has also shown mentions of possible mental disorders such as de\-pression\textsuperscript{\ref{trained-psych}}.

Unhappiness appears to also bring down motivation among developers, which is a critical force in software engineering activities \citep{Franca2014b}. Negative experiences and negative affect are also perceived to be causes of work withdrawal. Psychology research has recently started to investigate the role of affect in work withdrawal (e.g., \citet{Miner2010}), and software engineering research is lacking on the matter. Yet, negative affect has been found to be a predictor of, e.g., leaving Open Source projects, as observed in linguistic analysis of mailing list posts~\citep{rigby2007}. Our analysis shows that developers may distance themselves from the task to which their unhappiness relates, up to the point of quitting jobs.

Among the least mentioned categories, we wish to highlight reduced creativity as a consequence of unhappiness. Creativity has been considered beneficial and even required in software development (see, e.g., \citet{brooks1975}). Its reduction should be considered problematic in many situations.

\paragraph{Consequences of Happiness}

In our analysis for RQ2, we found 32 categories of positive consequences of happiness. Many of these have a positive impact on software engineering outcomes. Higher productivity and performance appear as aspects which benefit most from happiness. Approximately 33\% of the related in-text references, categorized into high cognitive performance and process-related productivity, deal with increased productivity and performance. This strengthens the result obtained for RQ1: developers experience happiness as a productivity and performance booster, and unhappiness has the opposite effect.

More than half of the consequences related to developers' experiences of happiness (58\%) had to do with the developer's own being. The most prevalent category is high cognitive performance. Happy developers appear to be more focused, have higher performance on problem-solving, higher mental energy, higher skills, and to learn better. Again, given the nature of the software development task, these consequences are very desirable, and can result in positive external outcomes.

Another prevalent category of consequences is high motivation, which participants reported experiencing due to happiness. This suggests that influencing developers' happiness should be considered when the goal is to influence their motivation. A perceived positive atmosphere, with peace of mind and opportunities to enjoy the moment, are factors contingent on the social environment. \citet{Fagerholm2015} has previously linked a positive atmosphere to high performance.

Other positive outcomes reported by our participants include high work engagement and perseverance, higher creativity, higher self-confidence, and being valued and proud.

Less than half of the consequences related to happiness (42\%) were external. As with unhappiness, process-related outcomes were most prevalent. Participants experienced high productivity, expediation, sustained flow, increased collaboration, increased process adherence, and a more creative process to result from happiness. This indicates a more speedy, even process with more social ties to other developers, and a stronger commitment to follow the agreed process, including ``doing things right'', writing documentation, following best practice, and writing tests.

A smaller but still meaningful portion of the external consequences had to do with the software artifact under development. Here, participants exclusively discussed high code quality as a consequence of happiness.

\paragraph{Comparing Consequences of Unhappiness and Happiness}

When comparing the results for the two research questions, some interesting insights emerge. Many of the consequences have similar prevalence on both the happiness and unhappiness side.

While our results show more consequences of unhappiness than consequences of happiness (42 vs.\ 32; see figures~\ref{fig:categories-unhappiness} and~\ref{fig:categories-happiness}), there are more coded instances regarding happiness (340) than unhappiness (250). In other words, participants have reported consequences of unhappiness in more detail, but they have reported more consequences of happiness in total. This is despite the fact that we attempted to produce two categorisation schemes that were comparable -- the data did not permit the two to be one-to-one mirror images of each other. Our explanation for this seemingly contradictory finding is that two factors could be in play simultaneously: It is known that humans are prone to focus on negative experiences more than positive ones (e.g.,~\citet{baumeister2001}). Thus, participants may be able to recall such experiences in greater detail. Simultaneously, software developers seem to be a slightly happy population (see \citet{Graziotin:2017ease}, where results are from the same sample as this study). Thus, participants may be able to recall more positive experiences in total over the entire set of responses. This finding underscores the need to focus on limiting unhappiness despite developers being happy overall.

Categories in the developer's own being have similar prevalence across unhappiness and happiness. Both in experienced consequences of unhappiness and happiness, cognitive performance is the most prevalent category, with unhappiness reducing it and happiness increasing it. The structure within those categories is also very similar for both sides, e.g. low focus is first on the unhappiness side, while being more focused is first on the happiness side. Higher problem-solving performance does not have a counterpart on the unhappiness side. This category could be related to being stuck in problem-solving, which we have previously reported as an important cause of unhappiness \citep{Graziotin:2017ease}.

The second most prevalent consequence of unhappiness is mental unease or disorder, with no direct counterpart on the happiness side. Several other consequences might be construed as mental well-being or positive mental health, but no single consequence can be said to be the mirror category.

Lower motivation is the third most prevalent consequence of unhappiness and its opposite, high motivation, the second most prevalent consequence of happiness. This highlights its importance in developers' experience of consequences of both happiness and unhappiness.

The fourth most prevalent consequence of unhappiness, and the fifth of happiness, are work withdrawal and high work engagement and perseverance. These categories are related to motivation, but the latter is more general.

The remaining positions in terms of prevalence are creativity (fifth, unhappiness, sixth, happiness) being cautious -- higher self-confidence (both on seventh place); undervalued (tenth, unhappiness) -- being valued (eighth, happiness); unhealthy coping behavior (eleventh, unhappiness) -- healthy coping behavior (tenth, happiness).

Higher self-accomplishment has no direct counterpart on the unhappiness side. This category pertains solely to achievement and having reached something. Higher self-confidence is a related category with no direct counterpart on the unhappiness side. It is less connected to achievement, and although the two may influence each other, they are separate. On the unhappiness side, self-doubt and being cautious are similar categories, but higher self-confidence is more specific: feeling safe and courageous to engage in new or risky actions.

Among the external categories, the similarities in prevalence is also visible. Both in experienced consequences of unhappiness and happiness, consequences for productivity are the most prevalent within the process category. The process category on both sides have similar frequencies (102 and 104, or 75\% and 73\% for unhappiness and happiness, respectively) and the top codes have a similar structure. One category on the happiness side without a counterpart on the unhappiness side is increased collaboration. Happiness could lead developers to reach out to others more.

Artifact and working with artifact have similar frequencies on both sides (38 and 35, or 27\% and 26\%, respectively). On the happiness side, there is only one strong sub-category. Software developers' experiences of consequences of (un)happiness related to the software artifact appears to be largely connected to code quality.

Finally, we observe that only 22\% of the respondents in this study are in both the data set for experienced consequences of happiness and unhappiness. This means that the responses on happiness are, to a large extent, distinct from the responses on unhappiness. There may be a tendency among some respondents to recall and report more on happiness, and among others to report more on unhappiness. If this is the case, personality may explain why some respondents would focus more on one or the other. Another possible explanation is respondent fatigue. However, judging from the overall responses, this does not seem to be the case, as most respondents who provided an answer to the first displayed open question also provided one for the second. The content of the answers is the largest determinant for whether a response was coded as expressing a consequence of happiness or unhappiness. We therefore suggest that personality should be taken into account when studying and considering (un)happiness among software developers.

\subsection{Limitations}
We elicited the experienced consequences of unhappiness and happiness of software developers using a survey approach and qualitative data analysis techniques. Whether causality can be inferred from research approaches other than controlled experiments, e.g., eliciting experiences from introspection in the context of qualitative research, is a matter of debate \citep{Creswell2009,Djamba:2002jo,Glaser:2013ha}. However, several authors, e.g., \citet{Glaser:2013ha}, take the stance that qualitative data analysis can be used to infer causality from the experience of \emph{human} participants, provided that there is a strong methodology for data gathering and analysis. In our case, we followed Grounded Theory coding methodology \citep{Corbin2008} in order to strengthen the validity of our results. Furthermore, our research goal was to elicit the consequences of unhappiness as experienced by developers themselves. As the consequences come from first-hand reports, and the analysis method rigorously assesses the expressed causal mechanism, we argue that they accurately represent the respondents' views. Still, it should be noted that the study does not verify whether participants can distinguish between affects and other experiences, nor whether they infer the correct direction of causality. Our study does not imply any general relationship between any specific consequences; only experienced consequences of (un)happiness are claimed.

Our sample of software developers using GitHub is limited in size and with respect to representativeness of developers at large. Our dataset (see Section \ref{sec:method:data-retrieval}) contains accounts with public activity during a six-month period. This may result in a bias as developers who prefer not to display their work in public would not be present in the data set; nor would developers whose work is done in companies' internal systems. The six-month time period, however, is less of an issue as very inactive developers are of less interest to our study -- but they may differ in terms of how they view consequences of (un)happiness. Replication using different data sources and collection methods is needed to validate our results in these scenarios.

As a small remark, we are aware that a random sample of GitHub projects is likely to bring noise in the data due to, for example, repositories containing only pictures, data, or even writings~\citep{Munaiah2017}. We believe that this issue, while very valid for most mining software repositories studies, does not pertain to the present study. We did not seek to sample projects but developers and their contact details. The contents of the sampled GitHub repository is not a concern for the present study. Furthermore, our invitation e-mail specified our quest for software developers only, and our questionnaire contained demographic items related to the participants' relationship with software development (e.g., developing software as volunteer/passion, as employee, or as freelancer; how much working time is dedicated to software development; and main role as software engineer).

Several potential threats to validity concern the study sample. Sampling from GitHub could lead to over-representation of open source developers in the sample through self-selection bias. We believe that GitHub is nowadays able to attract non-typical open source developers who are actively maintaining small projects while they are working for traditional companies. We show elsewhere~\citep{Graziotin:2017ease} that our sample was rich and able to capture a wide spectrum of participants: about 75\% of the respondents were professional software developers, 15\% were students, and 8\% were in other roles (such as manager, CEO, CTO, and academic researcher). The remaining participants were non-employed and not students. Whether any remaining bias in the sample has an impact on the experienced consequences of (un)happiness is an open question that requires replication with other samples to address. However, as can be seen in the results, there are many reported consequences that would not arise in simple, single-person projects, and several of the comments are made regarding a corporate environment. This indicates that results are relevant in professional settings.

It might be the case that the ``GitHub population of developers'' is slightly younger than developers in general, but to our knowledge no empirical evidence exists in either direction. GitHub is a reliable source for obtaining software engineering research data, as it allows replication of this study on the same or different populations. The GitHub community is large (30 million visitors per month~\citep{Doll:2015uv}) and diverse in terms terms of team size, type of software, and several other characteristics. Our sample is similarly diverse and is balanced in terms of demographic characteristics, including participant role, age, experience, work type, company size, and students versus workers. One exception is gender: our sample is strongly biased towards males. The direction of the bias may be the same as in the general population. Unfortunately, it is a known problem that software engineering roles are predominantly filled by males \citep{Ortu:2016gz,Terrell:2016dq,Ford:2016}, although recent research is attempting to tackle the issue. Data from some sources indicate between 7.6\%\footnote{StackOverflow Developer Survey 2017, \url{https://stackoverflow.com/insights/survey/2017}} and 20\%\footnote{Bureau of Labor Statistics Current Population Survey: Annual averages, Software developers, applications and systems software, \url{https://www.bls.gov/cps/cpsaat11.htm}} females, with numbers possibly depending on the definition of developer and the countries or cultures represented.

\subsection{Recommendations for Practitioners}

We believe that our discovered consequences of happiness and unhappiness of developers should be of interest to practitioners working as managers, team leaders, but also team members and solo software developers. To facilitate such use, we have made the category schemes available as archived open data \citep{Graziotin:2016ck}. Practitioners in leadership positions should attempt to foster overall happiness of software development teams by limiting their unhappiness and promoting factors that contribute to happy experiential episodes. The benefits of fostering happiness among developers were empirically demonstrated in past research, and they especially highlight software development productivity and software quality boosts. With our results, we add that addressing unhappiness will limit the damage in terms of several factors at the individual, artifact, and process level. We also add that addressing happiness can impact the very core prerequisites for software development: cognitive performance, motivation, and a positive atmosphere at the workplace. We note that previous research \citep{graziotin2015you} has suggested that intervening on the affect of developers might have relatively low costs and astonishing benefits.

\subsection{Implications for Researchers}

We believe that the results of the present work could be adopted as the basis of several research directions. Our study has the potential to open up new avenues in software engineering research based on the discovered factors (e.g., work withdrawal and affect of developers). Also, all the factors we have reported are the end part of an experienced causality chain with unhappiness or happiness as the antecedent. Future studies should attempt to seek a quantification of the chain.

We found that broadly speaking, developers more frequently consider their happiness to benefit themselves, and their unhappiness to be detrimental to others. The latter part of this observation was weaker than the former. Future research could attempt to uncover whether this finding holds in other samples of the software developer population, and, if so, investigate whether this is due to a bias among developers or if there are different underlying reasons.

Finally, as we have demonstrated, software developers experience a multitude of consequences arising from happiness and unhappiness. The consequences touch upon several issues that are of traditional interest in software engineering research, such as productivity in software processes, process adherence, and software quality. Our work indicates that (un)happiness, and, more generally, affect, should be taken into account in empirical studies investigating developers conducting activities related to such outcomes. This applies on the individual level, but also on aggregated levels of analysis, such as teams and organizations. We argue that ignoring the factors reported in this article can lead to incorrect inferences, or, at the very least, lead to omitting important confounding variables. Considering them could at best lead to novel insights and solution proposals for improving the practice of software engineering, and at least would give a voice to developers volunteering their time as research subjects. We call for software engineering researchers to take (un)happiness into account in their studies.

\section{Conclusion}

In this paper, we presented the results of an analysis of the experienced consequences of unhappiness and happiness among software developers while developing software. The complete results are archived and available as open data \citep{Graziotin:2016ck}. The consequences are grouped into the main categories of internal -- developer's own being -- and external -- process and artifact. The highest impact of both happiness and unhappiness is experienced to be on development productivity and quality as expressed by cognitive performance, including creativity and flow, and process-related performance. We found several instances of job-related adverse effects of unhappiness and even indications of mental disorders: work withdrawal, stress, anxiety, burnout, and depression\textsuperscript{\ref{trained-psych}}. We also found instances of effects beneficial to well-being, such as a perceived positive atmosphere at work, higher self-accomplishment, work engagement, perseverance, creativity, and self-confidence. Overall, our results indicate that developers experience happiness more as benefiting themselves, and unhappiness more as being detrimental to others.

Our recommendation to practitioners, including managers and team leaders, is to utilize our list of consequences and the explanations offered by the present paper to start their quest for enhancing the working conditions of software developers. The consequences, in particular, offer interesting angles which managers should reflect on and look out for in their workforce.

We believe that our study results are of immediate application in future academic work. The results set theoretical foundations for causality studies and inspiration for novel research activities in software engineering. Our own aims for future research include work towards completing the picture of (un)happiness among software developers in terms of causes, links between causes and consequences, and separating positive and negative causes and consequences of both happiness and unhappiness. Replication of the study both with other samples of the entire developer population as well as with samples of sub-populations such as programmers, tester, and architects, would also be of great value, and our work can be used to aid the design of such studies and as a basis for comparison.

The present study enforces the stance that many aspects of software engineering research require approaches from the behavioral and social sciences; we believe there is a need in future academic discussions to reflect on how software engineering research can be characterized in such terms. Developers are prone to share work-related horror stories on a daily basis, and we believe that their job conditions are often overlooked. With our past and present research activities, we hope we can contribute towards higher well-being of software engineers, while enhancing the amount and quality of their job outputs.

\section*{Acknowledgements}

The authors would like to thank all those who participated in this study. Daniel Graziotin has been supported by the Alexander von Humboldt (AvH) Foundation.

\section*{References}

\bibliography{references}

\end{document}